\documentclass{article}
\usepackage{spconf,amsmath,amssymb,graphicx,hyperref}

\title{CIF-QUERY BRIDGING FOR AUDIO-NATIVE ASR TRANSFER TO VISUAL SPEECH RECOGNITION}

\name{Chang Sun$^{1,2}$, Dongliang Xie$^{1}$, Wanpeng Xie$^{1}$, Bo Qin$^{2}$, Hong Yang$^{2}$}
\address{\small$^{1}$State Key Laboratory of Networking and Switching Technology, Beijing University of Posts and Telecommunications \\
\small$^{2}$First Research Institute of The Ministry of Public Security of PRC, Beijing 100048, China}

\begin{document}
\maketitle

\begin{abstract}
Audio-native automatic speech recognition (ASR) models provide strong
speech-language priors, but their compact acoustic interfaces are mismatched
with long, frame-level visual features. We propose \emph{CIF-Query Bridge}, a
lightweight connection from a Paraformer visual speech recognition (VSR)
encoder to Qwen3-ASR. The existing visual continuous integrate-and-fire (CIF)
predictor produces an input-dependent number of token-rate queries; a
two-layer cross-attention bridge refines them against the complete frame
sequence, and a linear adapter feeds the pretrained ASR projector and decoder.
This design preserves access to uncompressed visual context while avoiding
direct autoregressive processing of hundreds of video frames. On CMLR, the
complete visual-only transfer system achieves 7.97\% character error rate,
compared with 11.78\% for native Paraformer under the same test split and
scoring protocol---a 3.81-point absolute and 32.3\% relative reduction.
Controlled ablations characterize bridge depth, projector initialization,
adapter capacity, and CTC supervision. The results show that CIF-derived
adaptive queries provide a compact interface for transferring audio-native
ASR capability to VSR.
\end{abstract}

\begin{keywords}
visual speech recognition, lip reading, continuous integrate-and-fire,
audio-native ASR, cross-modal transfer
\end{keywords}

\begin{figure*}[!t]
  \centering
  \includegraphics[width=0.94\textwidth]{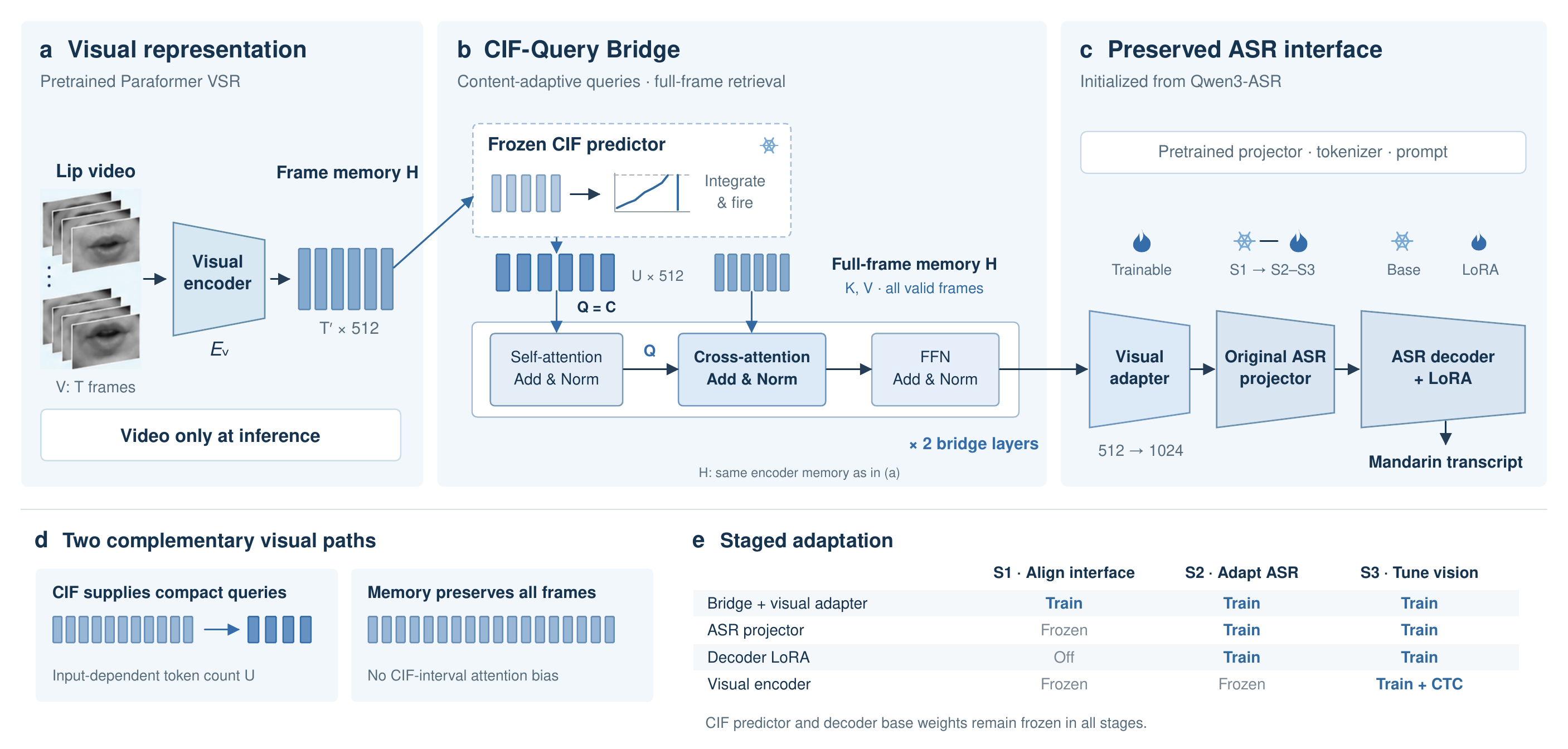}
  \caption{Overview of CIF-Query Bridge. CIF produces token-rate queries
  $\mathbf C$, while the complete visual sequence $\mathbf H$ remains
  cross-attention memory. The linear adapter is used in the main system; the
  residual variant is ablated. Module labels show the staged schedule.}
  \label{fig:architecture}
\end{figure*}

\section{Introduction}
\label{sec:intro}

Visual speech recognition (VSR), also known as lip reading, transcribes speech
from visible mouth motion alone. It is valuable when audio is unavailable or
corrupted and for silent or privacy-sensitive interaction. VSR is nevertheless
intrinsically ambiguous: distinct phonemes can share similar visemes, while
coarticulation makes a visual unit dependent on surrounding context. Early
end-to-end systems such as LipNet demonstrated joint visual feature extraction
and sentence prediction \cite{assael2016lipnet}. Mandarin presents an additional
challenge because visually similar pronunciations can correspond to different
characters and meanings. CMLR was introduced to study this large-vocabulary,
sentence-level setting \cite{zhao2019cmlr}.

Recent VSR advances combine stronger sequence modeling, multimodal pretraining,
and data scaling. Conformers established strong supervised baselines
\cite{ma2021conformer}; AV-HuBERT learns masked audio-visual units
\cite{shi2022avhubert}; VATLM unifies visual, audio, and text pretraining
\cite{zhu2024vatlm}; RAVEn transfers raw audio-video supervision
\cite{haliassos2023raven}; and Auto-AVSR scales training with automatic
transcripts \cite{ma2023autoavsr}. These methods learn strong visual encoders,
but their task-specific decoders provide less linguistic capacity for resolving
visual ambiguity.

LLMs offer stronger linguistic priors. VSP-LLM maps compressed visual speech
units to an LLM \cite{yeo2024vspllm}; VALLR reconstructs text from predicted
phonemes \cite{thomas2025vallr}, and VALLR-Pin adds Pinyin-guided Mandarin
refinement \cite{sun2025vallrpin}. Llama-AVSR uses modality projectors and LoRA
to connect audio-visual encoders to a generic LLM
\cite{cappellazzo2025llamaavsr,hu2022lora}, while MMS-LLaMA uses rate-aware
Q-Former queries \cite{yeo2025mmsllama}. Dynamic query allocation is therefore
not unique to our approach; here, an existing visual CIF predictor supplies
both query content and length. Unlike a generic text LLM, an audio-native ASR
model expects a compact sequence of acoustic representations. Frame-level
visual features differ in modality, dimension, and length, so the ASR decoder's
linguistic capability cannot be used directly without a dedicated connection.

We investigate how an audio-native ASR input interface can instead be preserved
and reused for visual-only recognition. Our \emph{CIF-Query Bridge} connects a
pretrained Paraformer VSR encoder to Qwen3-ASR \cite{shi2026qwen3asr}. We reuse,
rather than redesign, the VSR model's continuous integrate-and-fire predictor
\cite{dong2020cif,gao2022paraformer} to produce an input-dependent number of
token queries. Cross-attention refines these queries against the complete
frame-level visual memory before a small adapter and the original ASR projector.
This construction retains the speech interface and avoids presenting hundreds
of video frames directly to the autoregressive decoder.

Our contributions are threefold. First, we propose a compact CIF-Query Bridge
that connects a pretrained VSR front end to the Qwen3-ASR projector and
decoder. Second, we use existing CIF outputs as input-dependent queries that
can attend to the complete visual sequence, combining token-rate compression
with frame-level memory. Third, we report 7.97\% CER on CMLR and study bridge
depth, projector initialization, adapter capacity, and CTC supervision under
controlled schedules.

\section{Proposed Method}
\label{sec:method}

\subsection{Architecture overview}

As illustrated in Fig.~\ref{fig:architecture}, given a lip video $\mathbf{V}$
with $T$ frames, a pretrained visual encoder
$E_{\mathrm v}$ produces
\begin{equation}
 \mathbf{H}=E_{\mathrm v}(\mathbf{V})\in
 \mathbb{R}^{T'\times d_{\mathrm v}},
 \qquad d_{\mathrm v}=512,
\end{equation}
where padding positions are excluded by a binary mask. The system retains
$\mathbf H$ as frame-level memory. A frozen CIF predictor produces a shorter
sequence $\mathbf C$; a two-layer bridge uses $\mathbf C$ as queries and
$\mathbf H$ as keys and values. A visual adapter maps the bridge output to the
expected acoustic feature dimension. The resulting features pass through the
original Qwen3-ASR projector and replace the native speech placeholder in its
prompt. The tokenizer, prompt template, projector, and autoregressive decoder
therefore remain paired with their original ASR model. No audio or
audio-derived feature is required at inference.

\subsection{CIF-query cross-attention bridge}
\label{ssec:bridge}

We adopt the CIF predictor from the pretrained Paraformer VSR model without
changing its architecture. It estimates a weight $\alpha_t$ at each encoder
step and integrates consecutive weighted features until the accumulated mass
reaches one. This produces
\begin{equation}
 \begin{aligned}
 \mathbf C&=[\mathbf c_u]_{u=1}^{U}
   \in\mathbb{R}^{U\times d_{\mathrm v}},\\
 U&=\left\lfloor\sum_{t=1}^{T'}\alpha_t+\tau\right\rfloor.
 \end{aligned}
\end{equation}
Here $\tau=0.45$ is the inherited tail mass appended with a zero feature;
implementation guards clip the count to the available output length and
at least one position. Both adaptation and inference use predicted lengths,
without reference-length scaling. CIF supplies query content and count, not
Qwen text-token boundaries: the original predictor was trained with a
character-plus-EOS length target. Its outputs serve as queries with access to
uncompressed visual memory.

Let $\mathbf Z^0=\mathbf C$. Bridge layer $l$ applies token self-attention,
cross-attention to the visual memory, and a feed-forward network:
\begin{align}
 \widetilde{\mathbf Z}^{l} &={\rm LN}\left(\mathbf Z^{l-1}
   +{\rm MHSA}(\mathbf Z^{l-1})\right),\\
 \widehat{\mathbf Z}^{l} &={\rm LN}\left(\widetilde{\mathbf Z}^{l}
   +{\rm MHA}(\widetilde{\mathbf Z}^{l},\mathbf H,\mathbf H)\right),\\
 \mathbf Z^{l} &={\rm LN}\left(\widehat{\mathbf Z}^{l}
   +{\rm FFN}(\widehat{\mathbf Z}^{l})\right).
\end{align}
We use two layers, eight attention heads, GELU activations, and a fourfold FFN
expansion, with dropout 0.1 (omitted from the equations). We use no
CIF-interval distance bias. Predictor parameters remain frozen, but its
predicted integration weights and firing boundaries can change when the
visual encoder is updated.

\subsection{Visual-to-ASR interface projection}
\label{ssec:interface}

The bridge produces 512-dimensional visual tokens, whereas the Qwen3-ASR
projector expects 1024-dimensional acoustic features. We resolve this mismatch
with a visual adapter $G_{\mathrm v}$. Its minimal form is a linear map,
\begin{equation}
 \mathbf a=W_{\mathrm v}\mathbf Z^2+\mathbf b_{\mathrm v},
\end{equation}
followed by the retained two-layer ASR projector (biases omitted below)
\begin{equation}
 \mathbf A=W_2\,{\rm GELU}(W_1\mathbf a).
\end{equation}
The main system uses this linear adapter. The ablation also considers a gated
residual MLP ($1024\!\rightarrow\!2048\!\rightarrow\!1024$, dropout 0.1), whose
sigmoid gate is initialized with $g=-2$.

\subsection{Objective and staged adaptation}
\label{ssec:training}

The autoregressive decoder is optimized with teacher-forced cross-entropy
$\mathcal L_{\mathrm{AR}}$; prompt tokens and inserted visual positions are
masked from supervision. During final visual-encoder adaptation, its output
also receives CTC supervision:
\begin{equation}
 \mathcal L=\mathcal L_{\mathrm{AR}}+
 \lambda_{\mathrm{CTC}}\mathcal L_{\mathrm{CTC}},\qquad
 \lambda_{\mathrm{CTC}}=0.1.
\end{equation}
No CIF quantity loss is used. Freezing predictor parameters does not
guarantee invariant output lengths during visual-encoder adaptation.

Optimization proceeds in three stages. Stage~1 trains only the bridge and
visual adapter, while both pretrained backbones and the ASR projector are
frozen. Stage~2 additionally enables the original projector and rank-16 LoRA
modules in the language decoder. Stage~3 performs a short adaptation of all
visual encoder blocks at a much smaller learning rate and activates CTC. This
ordering first establishes the cross-modal interface before modifying either
pretrained backbone.

\section{Experiments}
\label{sec:experiments}

\subsection{Dataset and implementation}

We use the official CMLR split (61 hours, 11 speakers;
71,448/10,206/20,418 train/validation/test utterances) \cite{zhao2019cmlr},
grayscale $88\times88$ mouth crops, and CER. All our reported inputs are
visual-only. Corpus CER is the summed character edit distance divided by the
summed reference length; validation and test contain 145,832 and 292,180
reference characters. Following M2MeT, references are normalized for spaces,
annotations, punctuation, and Latin characters, while hypotheses only have
special tokens and ASR/thinking prefixes removed.

The 50-block Paraformer encoder/CIF predictor and Qwen3-ASR-1.7B initialize
the visual and language components. Training uses AdamW with zero weight
decay, bfloat16 on four GPUs, two-step gradient accumulation, clipping at 1.0,
cosine decay to 0.05 of the peak rate, and a 500-frame/GPU batch budget.
Stages~1--3 run for $8+4+1$ epochs. Stage~1 uses bridge/adapter LR
$2\times10^{-4}$ and 1,000 warm-up updates. Stage~2 uses LR
$2\times10^{-5}$ for the bridge, adapter, projector, and LoRA, with 400 warm-up
updates. Stage~3 uses LRs $10^{-7}$, $10^{-6}$, $10^{-5}$, and
$5\times10^{-6}$ for the encoder, CTC head, LoRA, and remaining trainable
modules, respectively, with 400 warm-up updates; it executes one epoch of an
eight-epoch cosine plan. Rank-16 LoRA uses scaling 32, dropout 0.05, and
q/k/v/o plus gate/up/down projections. Brightness, contrast, gamma, and
partition-mean temporal masking are applied only in Stage~3. Beam-8 decoding
generates at most 64 tokens, without sampling or CTC rescoring.

Controlled
variants use seed 42, the fixed schedule, and stage-end checkpoints. Validation
selects the two-layer linear model and chooses the random-projector Stage~1 LR
from $\{10^{-4},2\times10^{-4},5\times10^{-4}\}$; its later stages match the
pretrained-projector run. The no-CTC run branches from the residual model's
Stage~2 checkpoint.

\subsection{Comparison with prior work}

Table~\ref{tab:main} compares prior work with our native baseline.
``External'' denotes knowledge beyond the 61-hour CMLR training
set. Our visual encoder and CIF predictor use only CMLR. The only external
source is Qwen3-ASR, whose audio encoder is unused. The best listed result
using only CMLR is 9.10\% CER; 1,459 hours of AV pretraining gives 8.00\%.
Our 7.97\% is not a data-matched claim of superiority.

The native Paraformer baseline uses the same visual initialization checkpoint
with its original decoder, evaluated on the same test set with the same video
preprocessing and scoring protocol. It obtains 11.78\% CER with beam size one,
compared with 7.97\% for our transferred system with beam size eight. This is
an absolute reduction of 3.81 percentage points (approximately 32.3\% relative).
The comparison measures the complete transfer system, including decoder,
adaptation, and search changes; it does not isolate the CIF-query bridge.

\begin{table}[t]
\caption{Official CMLR test CER (\%); all systems use visual-only inference.
External: pretraining beyond CMLR; AV: audio-visual.}
\label{tab:main}
\centering
\vspace{4pt}
\setlength{\tabcolsep}{2.5pt}
\begin{tabular}{@{}lccr@{}}
\hline
Method & External & CMLR h & CER $\downarrow$ \\
\hline
CSSMCM \cite{zhao2019cmlr} & -- & 61 & 32.48 \\
LIBS \cite{zhao2020libs} & ASR teacher & 61 & 31.27 \\
CTCH \cite{ma2020ctch} & -- & 61 & 22.01 \\
JEP-KD \cite{sun2024jepkd} & ASR teacher & 61 & 14.26 \\
Ma et al. \cite{ma2022multilingual} & -- & 61 & 9.10 \\
Ma et al. \cite{ma2022multilingual} & 818-h AV & 61 & 8.10 \\
Ma et al. \cite{ma2022multilingual} & 1459-h AV & 61 & 8.00 \\
\hline
Native Paraformer (ours) & -- & 61 & 11.78 \\
\textbf{Ours (linear adapter)} & Qwen3-ASR & 61 & \textbf{7.97} \\
\hline
\end{tabular}
\end{table}

Qwen3-ASR reports 40 million hours for AuT and 3 trillion tokens for Omni
pretraining \cite{shi2026qwen3asr}. The stages are distinct, and their corpus
inventory and overlap with CMLR are unknown; our result therefore measures
foundation-model transfer rather than a data-matched comparison.

\subsection{Ablation study}
\label{ssec:ablation}

Table~\ref{tab:ablation} reports validation and test CER under the fixed
schedule. For a fair residual-model control, the random projector is trained
from the first Stage~1 update; validation selects $2\times10^{-4}$ from
$\{10^{-4},2\times10^{-4},5\times10^{-4}\}$, and Stages~2--3 match the
pretrained run. Pretraining lowers test CER from 8.2196\% to 7.9923\%
(paired 95\% CI $[-0.3270,-0.1282]$).

\begin{table}[t]
\caption{Controlled CMLR ablation, CER (\%). Proj.: projector initialization;
Res.: gated residual adapter. All rows use one training seed.}
\label{tab:ablation}
\centering
\vspace{4pt}
\setlength{\tabcolsep}{3pt}
\begin{tabular}{ccclrr}
\hline
Proj. & Layers & Adapter & CTC & Val & Test \\
\hline
Random & 2 & Res. & Yes & 8.1464 & 8.2196 \\
Pretr. & 0 & Linear & Yes & 8.1210 & 8.0820 \\
Pretr. & 1 & Linear & Yes & 8.0140 & 7.9971 \\
Pretr. & 2 & Linear & Yes & \textbf{7.8858} & \textbf{7.9721} \\
Pretr. & 2 & Res. & Yes & 8.0702 & 7.9923 \\
Pretr. & 2 & Res. & No & 8.1018 & 8.0776 \\
\hline
\end{tabular}
\end{table}

\begin{figure}[t]
  \centering
  \includegraphics[width=\columnwidth]{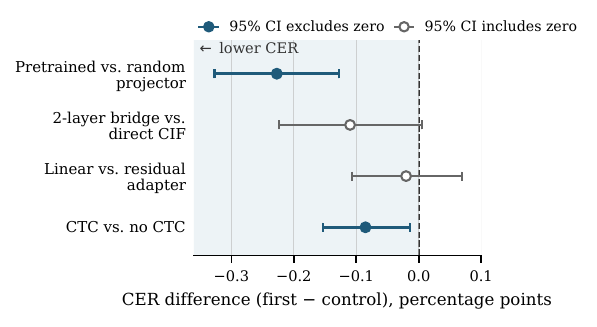}
  \caption{Paired test-CER differences with 95\% utterance-bootstrap CIs
  (10,000 samples); negative values favor the first configuration.}
  \label{fig:component_effects}
\end{figure}

With a linear adapter, depth 0-to-2 monotonically lowers validation/test CER
by 0.2352/0.1099 points; the test CI $[-0.223,+0.005]$ includes zero. Linear
versus residual has no clear test advantage ($-0.0202$;
$[-0.106,+0.069]$), whereas CTC lowers residual-model test CER by 0.0852
points ($[-0.154,-0.014]$). Figure~\ref{fig:component_effects} reports paired
utterance-bootstrap CIs for fixed models; they exclude seed variability, and
validation remains model-selection rather than independent evidence.

\section{Discussion}
\label{sec:discussion}

The 3.81-point complete-system gain likely reflects primarily the Qwen3-ASR
decoder and its linguistic prior. This does not make the connection secondary:
the ASR backend expects a short sequence of 1024-dimensional acoustic
representations, whereas the visual front end emits a much longer sequence of
512-dimensional features. Without adaptive rate compression, dimensional
alignment, and access to frame-level memory, the decoder's pretrained
capability cannot be applied to visual input. Our contribution is this
connection rather than the decoder itself. Moreover, the native comparison
changes the decoder, adaptation, and beam search, so it measures the complete
transfer system and does not isolate the bridge's causal contribution.

Within the controlled runs, the monotonic depth trend on both splits is
consistent with full-memory retrieval recovering context that token-rate CIF
integration may discard. The test interval nevertheless includes zero, and
depth also changes capacity and optimization. Equal-length pooled or learned
queries and a matched-capacity no-memory control are needed to identify the
source of the trend. Although CIF parameters remain frozen, Stage~3 changes
their encoder inputs and can shift firing weights, boundaries, and predicted
lengths; measuring this drift and its relation to CER is another useful test.

Evidence remains limited to one seed, CMLR, and one Paraformer--Qwen3-ASR
pairing. Historical test feedback weakens independent confirmation, and the
overlap between foundation-model pretraining and CMLR is unknown. Multiple
seeds, a matched text-LLM baseline, and independent datasets are needed to
establish stability and generalization.

\section{Conclusion}
\label{sec:conclusion}

We proposed CIF-Query Bridge, which uses adaptive CIF outputs as token-rate
queries over complete visual memory before the original audio-native ASR
interface. Combining content-adaptive compression with full-memory retrieval
avoids presenting one LLM token per video frame while retaining fine temporal
evidence. It also preserves the model-specific tokenizer, prompt, and
projector rather than learning an unrelated text-LLM input space. The resulting
visual-only Paraformer--Qwen3-ASR system achieves 7.97\% CER on CMLR, compared
with 11.78\% for native Paraformer under the same split and scoring protocol.
Controlled results show a consistent bridge-depth trend and delimit the
smaller effects of auxiliary components. They establish the feasibility of the
connection; future work should add multiple seeds, matched-query controls, and
evaluation on unseen datasets and languages.

\clearpage

\bibliographystyle{IEEEbib}
\bibliography{strings,refs}

\end{document}